\documentclass[a4paper]{mn2e}
\usepackage{natbib_jrm,times}
\usepackage{graphicx}
\usepackage{pdfpages}
\bibpunct[, ]{(}{)}{;}{a}{}{,}

\def \aj {AJ}
\def \mnras {MNRAS}
\def \pasp {PASP}
\def \apj {ApJ}
\def \apjs {ApJS}
\def \apjl {ApJL}
\def \aap {A\&A}
\def \nat {Nature}
\def \araa {ARAA}

\def\lesssim{\mathrel{\hbox{\rlap{\hbox{\lower4pt\hbox{$\sim$}}}\hbox{$<$}}}}
\def\gtrsim{\mathrel{\hbox{\rlap{\hbox{\lower4pt\hbox{$\sim$}}}\hbox{$>$}}}}

\long\def\symbolfootnote[#1]#2{\begingroup%
\def\thefootnote{\fnsymbol{footnote}}\footnote[#1]{#2}\endgroup} 

\begin{document}
\title[Spectropolarimetry of the Type IIb SN 2008aq]{Spectropolarimetry of the Type IIb SN 2008aq\thanks{Based on observations made with ESO Telescopes at the Paranal Observatory, under programme 080.D-0107}}

\author[Stevance et al.]{
\parbox[t]{\textwidth}{\raggedright
H.F. ~Stevance$^{1}$\thanks{f.stevance1@sheffield.ac.uk}, J.R. ~Maund$^{1}$\thanks{Royal Society Research Fellow}, D. ~Baade$^{2}$, P. ~H\"oflich$^{3}$, F. ~Patat$^{2}$, J. ~Spyromilio$^{2}$, \\ J.C. ~Wheeler$^{4}$,  A. Clocchiatti$^{5}$, L. ~Wang$^{6}$, Y. ~Yang$^{7}$, P. Zelaya$^{5}$}
\vspace*{6pt}\\
$^{1}$ University of Sheffield, Department of Physics and Astronomy, Hounsfield Rd, Sheffield S3 7RH, UK.\\
$^{2}$ European Organisation for Astronomical Research in the Southern Hemisphere, Karl-Schwarzschild-Str. 2, 85748 Garching. b. M\"unchen, Germany\\
$^{3}$ Department of Physics, Florida State University, 315 Keen Building, Tallahassee, FL 32306-4350, USA\\
$^{4}$ Department of Astronomy, University of Texas at Austin, Austin, TX 78712-1205, USA\\ 
$^{5}$ Departamento de Astronomia y Astrofisica, Pontificia Universidad Catolica Casilla 306, Santiago 22, Chile\\ 
$^{6}$ Department of Physics, Texas A\&M University, College Station, TX 77843-4242, USA\\
}
\maketitle
\begin{abstract}
We present optical spectroscopy and spectropolarimetry of the Type IIb SN 2008aq  16 days and 27 days post-explosion. The spectrum of SN 2008aq remained dominated by $\mathrm{H\alpha}$ P Cygni profile at both epochs, but showed a significant increase in the strength of the helium features, which is characteristic of the transition undergone by supernovae between Type IIb and Type Ib. Comparison of the spectra of SN 2008aq to other Type IIb SNe (SN 1993J, SN 2011dh, and SN 2008ax) at similar epochs revealed that the helium lines in SN 2008aq are much weaker, suggesting that its progenitor was stripped to a lesser degree. SN 2008aq also showed significant levels of continuum polarisation at $p_{\rm cont}$= 0.70 ($\pm$ 0.22) \%  in the first epoch, increasing to $p_{\rm cont}$=1.21 ($\pm$ 0.33) \% by the second epoch. Moreover, the presence of loops in the $q-u$ planes of $\mathrm{H\alpha}$ and He I in the second epoch suggests a departure from axial symmetry.
\end{abstract}


\section{Introduction}
\label{intro}

When massive stars (M$_{\rm ZAMS} >$ 8 M$_{\odot}$) have exhausted their fuel and die, they give rise to some of the most powerful explosions in the Universe: core-collapse supernovae (CCSNe). CCSNe are divided into a number of sub-classes: Type IIP/L SNe, which show prominent hydrogen lines; Type Ib/c SNe, which result from the explosion of progenitors that have been stripped of their hydrogen, or even helium layers by strong winds or binary interactions \citep{Filippenko97}; and Type IIb SNe, which, despite representing only a small fraction of CCSNe ($\sim$ 11\%), are an essential transitional class whose members evolve from Type II into Type Ib/c SNe. The progenitors of Type IIb SNe are stripped of nearly all of their hydrogen envelope, retaining less than 0.5 M$_{\odot}$ \citep{smith11}. As such, Type IIb SNe are sensitive probes of mass loss processes, particularly binary interactions (e.g \citealt{maund93J}, \citealt{fox14}).

Although we know about the progenitors of CCSNe (see \citealt{smartt09} for a review), the nature of the explosion mechanism remains a partial mystery. A variety of models have been proposed (for reviews in this field see \citealt{janka12} and \citealt{burrows13}), whose distinguishing observational features can be the geometries of the resulting ejecta. Spectropolarimetry is a unique tool that allows us to probe the 3D shapes of the ejecta of distant  SNe at early times. 

Linear polarization of the light emitted by SNe is the result of electron scattering, which is the principal source of opacity at early times \citep{ss82}. The polarization vector of a scattered photon will be perpendicular to the plane of scattering, defined as the plane containing the incident and scattered ray.  Consequently, in a spatially unresolved spherical envelope, the polarization vectors originated from regions located $\pi$/2 away from each other will cancel out, resulting in zero net polarization. A departure from spherical symmetry, however, will result in incomplete cancellation, and a polarization excess \citep{ss82,mccall84}, which can then be detected and used to quantify the shape of the ejecta. The intrinsic polarization associated with the continuum is closely related to the ellipticity of the envelope \citep{hoflich91}, while the polarization associated with individual spectral lines probes structural asymmetries to smaller scales \citep{WW08}.

Virtually all CCSNe show an excess of polarization \citep{WW08}, but the extent of that excess varies greatly from one type to another. Generally, Type II SNe show low polarization at early times, but as their envelope expands, and the inner core (beneath the hydrogen envelope) is revealed, the levels of polarization rise significantly (e.g.  \citealt{leonard06}). Type Ib/c SNe, on the other hand, tend to be significantly polarized at all epochs, and to a higher degree than Type IIP SNe (e.g. \citealt{maund08D}, \citealt{tanaka09} and \citealt{maund06aj}). Some Type IIb SNe exhibit early polarization levels similar to that of Type IIP/L SNe -- e.g SN 2001ig \citep{maund01ig} -- but more typical cases such as SN 1993J, SN 1996cb, SN 2008ax, or SN 2011dh show continuum polarization around $p \sim$ 0.5 \% to $p \sim$ 1 \% \citep{trammell93,wang01,chornock11,silverman09,mauerhan15}.

\begin{figure}
\centering
\includegraphics[width=6cm]{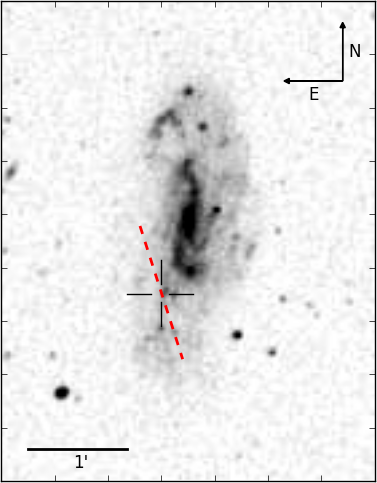}
\caption{SN 2008aq in MCG-02-33-20.The diagonal line represents the orientation of the interstellar polarisation (ISP), see section \ref{sec:isp}. This is a Digitized Sky Survey image, retrieved via Aladin.}
\label{fig:08aq}
\end{figure}

Here we report spectropolarimetric observations of the Type IIb SN 2008aq. SN 2008aq was discovered by \cite{08aq}, on February 27.44 2008, in the galaxy MCG-02-33-20 (see Figure \ref{fig:08aq}), which has a recessional velocity \footnote{Found on https://ned.ipac.caltech.edu} of 2407 km\,s$^{-1}$, and was subsequently classified by \cite{modjaz14} as a Type IIb SN. Based on a comparison of the lightcurve of SN 2008aq published by \cite{bianco14} to that of similar Type IIb SNe (SN 1993J and SN 2008ax), we estimate that SN 2008aq was discovered approximately 8 days before V-band maximum. The date of the explosion was taken to be 20 days prior to V-band maximum, as done by \cite{kumar13}, following the study of Type IIb and Type Ib/c SNe lightcurves by \cite{richardson06}. Consequently, we conclude that SN 2008aq exploded on February 16 2008.

\section{Observations and Data Reduction}
\label{sec:obs}

Spectropolarimetric observations of SN 2008aq were acquired with Focal Reducer and low dispersion Spectrograph (FORS) on the European Southern Observatory (ESO Very Large Telescope (VLT), in its dual-beam spectropolarimeter “PMOS” mode \citep{1998Msngr..94....1A}.  Observations of SN 2008aq were conducted at two epochs: 2008 Mar 4.3 and 15.2, a few days before and about a week after V-Band maximum, respectively. Based on our estimate of the explosion date, the observations correspond to $\sim$16 and 27 days post explosion. Both sets of observations used the $300V$ grism, providing a spectral resolution of 12.5\r{A} at 6000\r{A} (as determined from arc lamp calibration frames).  These observations did not use an order separation filter, such that the observations covered a wavelength of 3400-9300\r{A} at the expense, however, of possible second order contamination at redder wavelengths.  
The data were reduced in the standard manner using IRAF\footnote{IRAF is distributed by the National Optical Astronomy Observatory, which is operated by the Association of Universities for Research in Astronomy (AURA) under a cooperative agreement with the National Science Foundation.} following the prescription of \citet{maund05bf}.  The Stokes parameters were calculated following the  routines of \citet{2006PASP..118..146P}, with the data rebinned to 15\r{A} to improve levels of signal-to-noise.  We have also checked using standard stars observed with and without order sorting filters, and the second order effects at 7000\r{A} (the cut-on wavelength of the grism efficiency) and long-wards are well below the noise level in our data.
Flux spectra of SN 2008aq were calibrated against observations of flux standard stars acquired with the polarimetry optics in place. 

\begin{table}
\caption{\label{tab:obs} VLT Observations of SN~2008aq.}
\begin{tabular}{lcccc}
\hline\hline
Object    &  Date   & Exposure   & Epoch  & Airmass \\
               &(UT)     & (s)              &  (d)       & (Avg.)   \\
\hline
SN 2008aq  & 2008 Mar 04.3 & $4\times 900$   & +16 & 1.119\\
EG274        & 2008 Mar 04.4   & $10$     & +16 & 1.047\\
\\
SN 2008aq & 2008 Mar 15.2 & $2 \times 4 \times 900$  & +27 & 1.073\\
GD108        & 2008 Mar 15.2   & 60 & +27 & 1.067\\
\hline\hline               
\end{tabular}
\end{table}

\section{Results}
\label{sec:res} 

\subsection{Optical Spectroscopy}

\begin{figure}
\centering
\includegraphics[width=8cm]{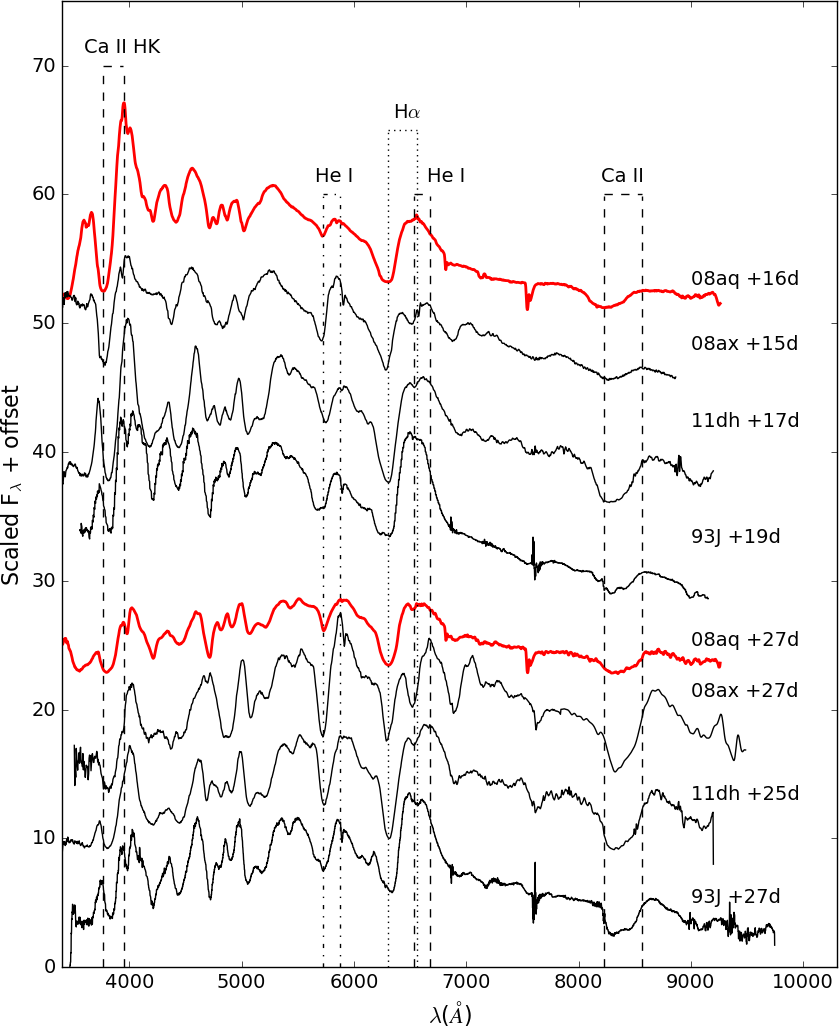}
\caption{Flux spectra of SN 2008aq (shown in red) at 16 and 27 days post-explosion. Also shown are spectra of SN 1993J, SN 2008ax, and SN 2011dh (in black). For each identified element, two lines were drawn, showing the rest wavelength and the wavelength at absorption minimum. The lines were identified by eye. All of the spectra are corrected for the recessional velocity of their respective host galaxies.}
\label{fig:comp}
\end{figure}

The flux spectra of SN 2008aq on 4 and 15 March 2008 are plotted in Figure \ref{fig:comp}, along with the flux spectra of SN 1993J, SN 2008ax and SN 2011dh (all type IIb SNe; obtained from WISeREP\footnote{http://wiserep.weizmann.ac.il/}) at similar epochs, for comparison.

At 16 days post-explosion, the spectrum of SN 2008aq  is dominated by broad P Cygni profiles of Ca II H\&K and $\mathrm{H\alpha}$; their absorption minima correspond to velocities of -13,700 km\,s$^{-1}$ and -12,000 km\,s$^{-1}$, respectively. The $\mathrm{H\alpha}$ emission appears to be flat topped, due to a weak blue-shifted He I $\lambda$6678 absorption feature, which is common to the other Type IIb SNe.  This feature evolves into a 'notch' at 6530 \r{A} by the second epoch. The velocity at the absorption minimum of $\mathrm{H\alpha}$ was found to be  -11,700 km\,s$^{-1}$ at +27 days. An absorption feature due to He I $\lambda$5876 was observed with a velocity of -7800 km\,s$^{-1}$ and -7350 km\,s$^{-1}$ in the first and second epochs, respectively. The strength of this feature increases by the second epoch, and a He I $\lambda$7065 feature emerges, with a velocity of -7000 km\,s$^{-1}$. Additionally, two narrow absorption lines can be seen superposed onto the emission component of He I $\lambda$5876, due to Na I D originating in the Milky Way, and at the recessional velocity of the host galaxy. The absorption observed around 8200 \r{A} is attributed to the Ca II Infra-Red (IR) triplet , with a velocity of -12,000 km\,s$^{-1}$ at 16 days post-explosion. The Ca II IR triplet P Cygni profile is peculiar in that it exhibits a relatively deep absorption paired with a weak emission component. \cite{branch02} noted that this behaviour is characteristic of a "detached" line forming region at velocities significantly greater than that of the photosphere. By the second epoch, 10 days later, the velocity of the Ca II IR triplet has decreased dramatically to a value of -9000 km\,s$^{-1}$ and the P Cygni profile shows a prominent emission component.
In the second epoch, the velocity of the Ca II IR triplet and Ca H \& K features has decreased more noticeably than that of other elements. 

The most remarkable difference between SN 2008aq and the comparison SNe (see Figure \ref{fig:comp}) is the relative weakness of the He I features. In the first epoch (+16 days) the He I $\lambda$5876 line is much shallower in SN 2008aq than in SN 2008ax, SN 1993J or SN 2011dh. Also the presence of a He I $\lambda$6678 feature can be inferred from the flat-topped profile of the $\mathrm{H\alpha}$ emission component, whereas in SN 2008ax, SN 2011dh and SN 1993J, this feature is clearly visible in the form of a "notch." In the second epoch (27 days post-explosion), the He I lines of SN 2008aq have significantly increased in strength, but are still much weaker than in the spectra of the other Type IIb SNe, and $\mathrm{H\alpha}$ remains the dominant feature. Additionally, the Ca II IR triplet feature at the second epoch is much shallower in the spectrum of SN 2008aq than in the spectra of 1993J, 2011dh and 2008ax, whereas in the first epoch the strength of that feature is similar for SN 2008aq, SN 2008ax and SN 2011dh. 


\subsection{Spectropolarimetry}
\label{sec:specpol}
\begin{figure*}
\centering
\includegraphics[width=15cm]{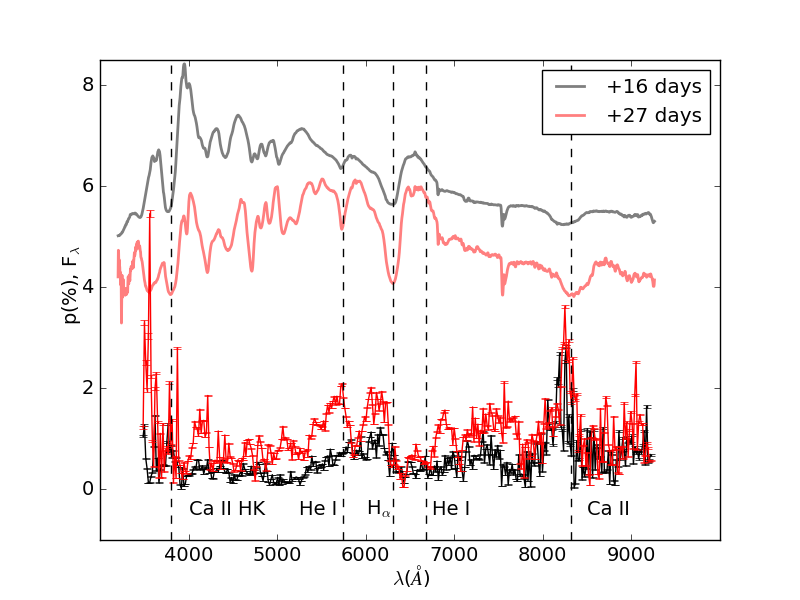}
\caption{Scaled Flux spectra (top) and polarization spectra (bottom) of SN 2008aq at 16 days and 27 days post-explosion. The polarization spectra were not corrected for interstellar polarization.}
\label{fig:pol}
\end{figure*}
The polarization and flux spectra for SN 2008aq are shown in Figure \ref{fig:pol}. The median polarization across the spectrum is $p \sim$ 0.5 \% in the first epoch, and rises to p $\sim$ 1 \% by the second epoch. Significant deviations from these values correlate with the major spectral features. The polarisation spectrum at 16 days is dominated by the broad inverse P Cygni profiles of $\mathrm{H\alpha}$ and the Ca II IR triplet, with maximum polarizations of 0.9 ($\pm$ 0.2)\% and 1.5 ($\pm$ 0.6) \%, respectively. Values of peak polarization were obtained by averaging the data in a range of 250 \r{A} around the centre of the peaks. It should be noted that the red part of the spectrum is contaminated by a high level of noise caused by fringing. At 27 days, the relative strength of the polarization associated with the He I $\lambda \lambda$ 5876, 6678, 7065 features has considerably grown, and the inverse P Cygni profile related to the He I $\lambda$5876 line is as prominent as that of $\mathrm{H\alpha}$. The polarization associated with the Ca II IR triplet has also increased by the second epoch, with a maximum around 2.8 ($\pm$ 0.4) \%. Also, depolarization is observed at both epochs between $\sim$ 4250 \r{A} to 5300 \r{A}, due to line blanketing caused by a blend of iron lines in that region of the spectrum.

\begin{figure}
\centering
\includegraphics[width=8cm]{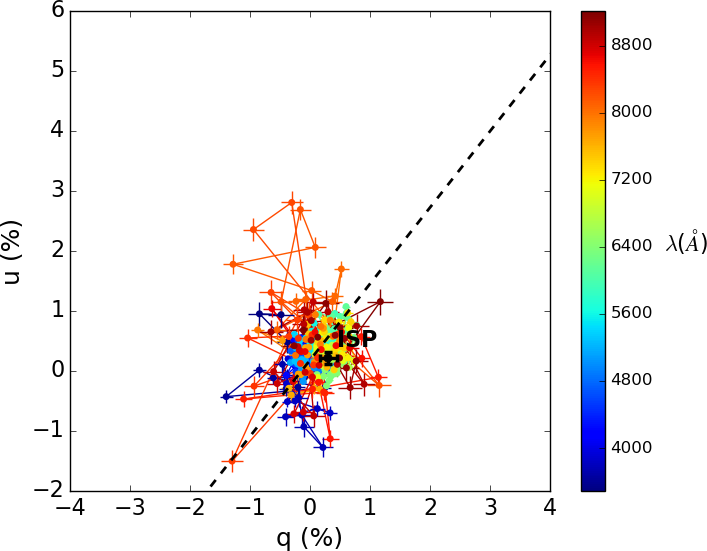}
\caption{Spectropolarimetric data of SN 2008aq on the 2008-03-04 (16 days after explosion), presented on the Stokes $q-u$ plane. The ISP is marked as a black point, and the dominant axis is represented by the dashed line. The data is colour coded according to wavelength.}
\label{fig:qu1}
\end{figure}

\begin{figure}
\centering
\includegraphics[width=8cm]{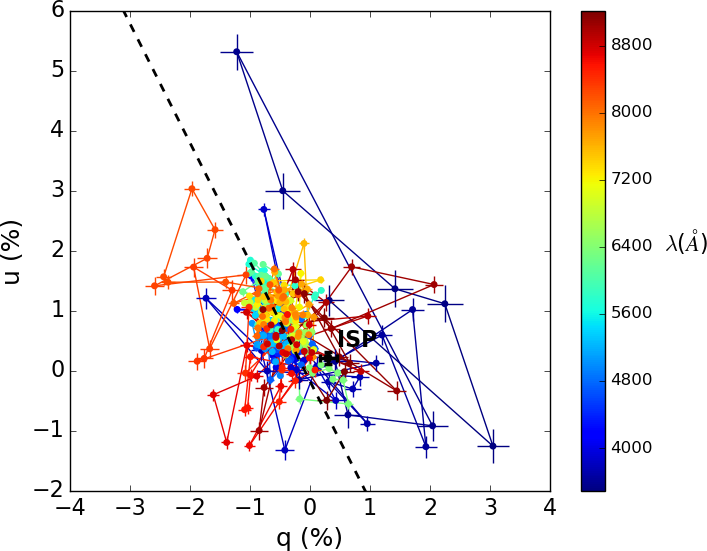}
\caption{Same figure as Figure \ref{fig:qu1}, but for the 15th of March 2008, or 27 days after explosion.}
\label{fig:qu2}
\end{figure}

The spectropolarimetric data at both epochs are plotted on the Stokes $q-u$ plane in Figures \ref{fig:qu1} and \ref{fig:qu2}. At both epochs, a dominant axis was fitted to the data using an Orthogonal Distance Regression package in python\footnote{http://docs.scipy.org/doc/scipy/reference/odr.html}. The whole spectrum lies along the respective dominant axes in the form of a tight, elongated cluster of points. The exception is the wavelength range associated with the Ca II IR triplet, which is clearly separate from the dominant axes and the rest of the data in both epochs. Fringing in this wavelength range results in a high level of noise on individual points, but the overall deviation of the Ca II IR triplet is real. A rotation of the dominant axis by $\sim$ 65 $^{\circ}$ is observed between both epochs, which corresponds to a change in P.A $\sim$ 32.5 $^{\circ}$.

\section{Analysis}
\label{sec:ana}
\subsection{Interstellar Polarisation}
\label{sec:isp}
The light we receive from the SN may also be polarised by dust in the interstellar medium between us and the object. The interstellar polarization (ISP) component must be quantified to permit proper analysis of the data. 

If the assumption of a standard Serkowski-Galactic type ISP is made \citep{serkowski73}, the ISP can be constrained using the relationship $p_{\rm ISP} \leq 9 \times E(B-V)_{\rm total}$, where $E(B-V)_{\rm total}$ is the sum of the reddening in the Milky Way and the host galaxy towards SN 2008aq. An empirical relationship relating the equivalent width of the sodium lines and the reddening was derived by \cite{poznanski12}. From the NaI D of the Milky Way component, we estimate that the reddening associated with dust in our Galaxy is $E(B-V)_{\rm MW}$ = 0.045 mag, which is in agreement with the estimates of foreground reddening of $E(B-V)_{\rm MW}$ = 0.04 mag by \cite{schlafly11}\footnote{https://ned.ipac.caltech.edu/}. The reddening associated with the host galaxy was estimated from the corresponding Na I D line, yielding $E(B-V)_{\rm host}$ = 0.027 mag. Consequently, the total reddening, is $E(B-V)_{\rm total}$ = 0.072 mag, which is similar to the value found and used by \cite{stritzinger09}. The upper limit on the ISP associated with our data is therefore 0.65\%.

If we make the assumption that the emission component of $\mathrm{H\alpha}$ is intrinsically unpolarised at early times \citep{tran97}, then the average polarization over that feature must be due solely to the ISP -- for a discussion on the veracity of this assumption see Section \ref{sec:disc}. Therefore we averaged the values of the Stokes parameters in the range 6700-6900 \r{A} to determine more precise values of the ISP, and found  $q_{\rm ISP}$ = 0.31 ($\pm$ 0.14) and $u_{\rm ISP}$ = 0.22 ($\pm$ 0.10), corresponding to $p_{\rm ISP} \sim$ 0.38\% . These values were subsequently used for the correction of the ISP.

\subsection{Continuum polarization}
\label{sec:cont}
After correction for the ISP, the polarisation level of the continuum was calculated by averaging the values of $p$ in the range 7000-7500\r{A}. We assumed the range 7000-7500\r{A} was representative of the continuum due to the absence of any strong lines in the flux spectrum. The corresponding uncertainty was taken to be the standard deviation of the polarization over this range. Values of $p_{\rm cont}$= 0.70 ($\pm$ 0.22) \% and $p_{\rm cont}$=1.21 ($\pm$ 0.33) \% were measured at +16 days and +27 days, respectively. The polarisation of the continuum is an indication of the overall geometry of the envelope. \cite{hoflich91} performed  Monte Carlo calculations for axi-symmetric scattering dominated atmospheres, and his Figure 4 shows the relation between the continuum polarisation and the axis ratio of the envelope. Comparing our values of the continuum polarisation to \cite{hoflich91}, the axis ratios at the first and the second epoch were $\sim$ 0.8-0.9, under the assumption that the continuum polarization is solely due to the geometry of the ejecta.

The polarisation angle of the continuum was evaluated in the same wavelength range as $p_{\rm cont}$, and found to be 55$^{\circ}$ ($\pm$ 44$^{\circ}$) in the first epoch, and 67$^{\circ}$($\pm$ 20$^{\circ}$) in the second epoch. Within the given uncertainties, it is unclear if the continuum polarization angle remained fixed between 16 days and 27 days post-explosion, or underwent a change similar to the rotation of the dominant axis ($\sim$ 32.5$^{\circ}$), see Section \ref{sec:specpol}.

\subsection{Line polarization}
\label{sec:line}
\begin{figure*}
\centering
\includegraphics[width=15cm]{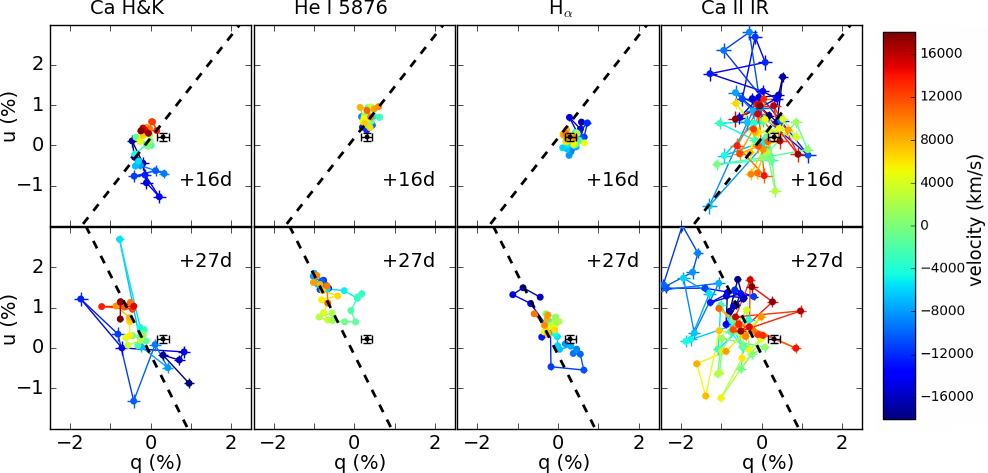}
\caption{Spectropolarimetric data associated with the absorption components of the P Cygni profiles of Ca H\&K, $\mathrm{H\alpha}$, He I $\lambda$5876 and the Ca II IR triplet at +16 days (top panels) and +27 days (bottom panels). The color scheme illustrates their velocity with respect to rest wavelength. The black point represents the ISP, and the dashed lines represent the same dominant axes as in Figures \ref{fig:qu1} and \ref{fig:qu2}. }
\label{fig:qu_line}
\end{figure*}
The polarization measured across Ca H\& K,  He I $\lambda$5876, $\mathrm{H\alpha}$ and the Ca II IR triplet is presented in the form of $q-u$ plots (see Figure \ref{fig:qu_line}).

At +16 days, He I $\lambda$5876 and $\mathrm{H\alpha}$ show low intrinsic polarization levels (i.e $q$ and $u$ close to 0), and the data are clustered close to the dominant axis. By + 27 days, clear loops arise, both in He I $\lambda$5876 and $\mathrm{H\alpha}$, which follow the direction of the dominant axis. Loops on the $q-u$ plane indicate that the degree of polarization changes with velocity, and hence with depth, suggesting departure from axi-symmetry on a small scale. The superposition of loops from line polarization along a common direction, however, is principally responsible for the dominant axes as they appear on Figures \ref{fig:qu1} and \ref{fig:qu2}. The rotation of the dominant axes between the two epochs implies a real rotation in the axial symmetry of the ejecta (in particular line forming region), even though a change in the continuum polarization is inconclusive (see Section \ref{sec:cont}).

The data associated with Ca H\& K 16 days post-explosion also show a loop, but it is not oriented in the direction of the dominant axis, which could suggest a significant departure from axial symmetry.  At +27 days, the data show substantial scatter, which is most likely the result of the high level of noise in this region of the spectrum, therefore making definite conclusions challenging to draw. The Ca II IR triplet data approximately follow the dominant axes at both epochs, but the signal is also very noisy and it is not possible to confidently ascertain the presence of loops.

\section{Discussion \& Conclusion}
\label{sec:disc}
 
Optical spectra and spectropolarimetric data of SN 2008aq were presented at 2 epochs: 16 days and 27 days post-explosion. The intrinsic polarization calculated for SN 2008aq at +16 days ($p_{\rm cont}$= 0.70 $\pm$ 0.22 \%) is similar to that of SN 2008ax at +9 days ($p_{\rm cont}$= 0.64 $\pm$ 0.02 \%) and SN 2011dh at +14 days ($p_{\rm cont} \sim$ 0.5 \%) \citep{chornock11,mauerhan15}. By the second epoch, the continuum polarization of SN 2008aq had reached a value of $p_{\rm cont}$=1.21 ($\pm$ 0.33) \%, which is close to the continuum polarization calculated for SN 1993J at +29 days ($p_{\rm cont} \sim$ 1 \%) by \cite{tran97}. 

A characteristic of Type IIb SNe is the increase in strength of their He features with time, as they transition to Type Ib SNe. This behaviour is observed in SN 2008aq (see Section 3.1), however, comparison with other Type IIb SNe at similar epochs reveals that the He features in the spectra of SN 2008aq are significantly weaker (see Figure 2). Additionally, we found that the pseudo equivalent width of the He I $\lambda$5876 absorption component at both epochs was roughly 4 to 5 times smaller in SN 2008aq than in other Type IIb SNe \citep{liu15}. The scarcity of helium indicates that the receding photosphere of SN 2008aq reached the helium layer at a later date than in SN2008ax, SN 2011dh, SN 1993J, and other previously studied Type IIb SNe; this may suggest that the progenitor of SN 2008aq was stripped off hydrogen to a lesser extent than other Type IIb SNe.

Like other Type IIb SNe -- e.g SN 1993J \citep{tran97} and SN 2001ig \citep{maund01ig}, SN 2008aq showed a drastic dominant axis rotation between +16 and +27 days (see Figures \ref{fig:qu1} and \ref{fig:qu2}), suggesting a change in the axis of symmetry deeper into the envelope. Additionally, as we have seen in Section \ref{sec:line}, loops arise in the $q-u$ planes of $\mathrm{H\alpha}$ and He I $\lambda$5876 by +27 days, indicating departures from axial symmetry in the $\mathrm{H\alpha}$ and He I line forming regions \citep{WW08}. Therefore, both the overall geometry and smaller scale geometry of the envelope of SN 2008aq vary with depth. Additionally, the polarization of the Ca II IR triplet is greater than that of $\mathrm{H\alpha}$ and He I at both epochs. Also considering the dramatic decrease in velocity of Ca II between the two epochs, compared to $\mathrm{H\alpha}$ and He I, it can be concluded that the line forming region of Ca II and the line forming region of $\mathrm{H\alpha}$ and HeI are separate.

Our determination of the ISP is dependent on the assumption that the Halpha P Cygni profile at early times is completely  depolarised (e.g \citealt{trammell93}, \citealt{tran97}), however, the emission component of $\mathrm{H\alpha}$ at +16 days is flat-topped which, as mentioned in Section \ref{sec:res}, is indicative of the presence of a He I $\lambda$6678 feature. Hence, the polarization associated with this part of the $\mathrm{H\alpha}$ P Cygni profile might be contaminated by polarization associated with He I $\lambda$6678. If, as we have shown in \cite{maund01ig}, contamination by the HeI 6678 feature masks intrinsic polarisation then our determination of the ISP is likely to be an underestimate. Since this work studies the variations of the polarisation across the lines, the absolute value of the derived polarisation is of interest but not critical to our conclusions. It is also worth noting that the ISP has a wavelength dependence \citep{serkowski73}. The ISP we derive, however, is small and therefore the wavelength dependence is likely to be negligible for our results.

\section*{Acknowledgments} 
The authors would like to thank ESO for the generous allocation of observing time, as well as the staff of the Paranal Observatory for their work on the program 080.D-0107. FS is supported through a PhD scholarship granted by the University of Sheffield. The research of JRM is supported through a Royal Society University Research Fellowship. JCW is supported by NSF Grant AST 11-09881.

\bibliographystyle{mn2e}

\end{document}